\documentclass[%
 reprint,
 amsmath,amssymb,
 aps,
aas_macros]{revtex4-2}

\usepackage{graphicx}
\usepackage{dcolumn}
\usepackage{bm}
\usepackage{ulem}[normalem]
\usepackage{xcolor}
\usepackage{amssymb}
\usepackage{aas_macros}
\usepackage{hyperref}

\begin{document}




\title{Variable gravitational potential of Milky Way analogues in HESTIA suite}






\author{Naira R. Arakelyan}
\affiliation{Astro Space Center of P.N. Lebedev Physical Institute, Moscow, Russia}

\author{Sergey V. Pilipenko}
\email{spilipenko@asc.rssi.ru}
\affiliation{Astro Space Center of P.N. Lebedev Physical Institute, Moscow, Russia}

\author{Stefan Gottl\"{o}ber}
\affiliation{Leibniz-Institut f\"{u}r Astrophysik Potsdam (AIP), An der Sternwarte 16, D-14482 Potsdam, Germany}


\author{Noam I. Libeskind}
\affiliation{Leibniz-Institut f\"{u}r Astrophysik Potsdam (AIP), An der Sternwarte 16, D-14482 Potsdam, Germany}

\author{Gustavo Yepes}
\affiliation{Departamento de F\'{i}sica Te\'{o}rica, M\'{o}dulo 15, Facultad de Ciencias, Universidad Aut\'{o}noma de Madrid, 28049 Madrid, Spain}
\affiliation{Centro de Investigaci\'{o}n Avanzada en F\'{i}sica Fundamental (CIAFF), Facultad de Ciencias, Universidad Aut\'{o}noma de Madrid, 28049 Madrid, Spain}

\author{Yehuda Hoffman}
\affiliation{Racah Institute of Physics, Hebrew University, Jerusalem 91904, Israel}

\date{\today}

\begin{abstract}
Investigations of trajectories of various objects orbiting the Milky Way (MW) halo with modern precision, achievable in observations by Gaia, requires sophisticated, non-stationary models of the Galactic potential. In this paper we analyze the evolution of the spherical harmonics expansion of MW analogues potential in constrained simulations of the Local Group (LG) from the HESTIA suite. We find that at distances $r\ge 100$~kpc the non-spherical part of the potential demonstrates a significant impact of the environment: ignoring the mass distribution outside the virial radius of the MW results in $>$20\% errors in the potential quadrupole at these distances. {Account of the environment results in a noticeable change of the angular momenta of objects orbiting MW analogues}. Spherical harmonics vary significantly during the last 6 Gyr. We attribute variations of the potential at $r\ge 30$~kpc to the motions of MW satellites and LG galaxies. We also predict that the non-sphericity of the real MW potential should grow with distance in the range $r_\mathrm{vir}<r<500$~kpc, since all realizations of simulated MW-like objects demonstrate such a trend.
\end{abstract}

\maketitle






\section{Introduction}
\label{sec:introduction}
Knowing the potential of the Galaxy is important for studying many aspects of astrophysics and cosmology. Several static models of the potential are widely used in the literature \cite{Ibata01,Helmi04,Law05,LM10,Koposov10,Kupper15,Malhan19}. However \cite{DSouza22,CorreaMagnus22} have shown that simple time-independent potentials are not adequate to trace back the orbital evolution of substructures inside the Milky Way (MW). Since the MW contains a relatively large satellite, the Large Magellanic Cloud (LMC), one improvement to the model of the Galactic potential is to introduce a moving LMC analogue to a model with a static potential. This approach has been used in, e.g. \cite{CorreaMagnus22,Koposov23}.   

It is known that the LMC also significantly disturbs the Milky Way's halo \cite{2019ApJ...884...51G,2020ApJ...898....4C,2020MNRAS.498.5574E,2020MNRAS.494L..11P,2021NatAs...5..251P,2021ApJ...919..109G,2023Galax..11...59V,2024MNRAS.527..437V}. These perturbations of the dark matter (DM) halo are believed to develop over time and depend on distance \cite{2021ApJ...919..109G}, so they could not be described by a static model plus moving LMC, and further model improvements are required. In \cite{Makarov23} a live N-body model of the MW and LMC has been used to show that the inclusion of a moving LMC can solve the problem of the bulk motion of satellite galaxies relative to the Sun.  

%


On the other hand, cosmological simulations show that the gravitational potentials of galaxies can be even much more complex. In the hierarchical model of galaxy formation, the accretion of matter is driven by large-scale streams and this contributes to the non-spherical density distribution and angular momentum in the forming halo. Simulations within the $\Lambda$CDM paradigm predict that $\sim$90\% of dark matter halos are significantly triaxial and have measurable figure rotation\footnote{Figure rotation is the rotation of the shape of the potential. It is not necessarily connected with the rotation of particles constituting a halo.}  \cite{1992ApJ...401..441D,2004ApJ...616...27B,2007MNRAS.380..657B}.
Dissipative collapse of cold gas and the formation of stellar disks change the shape of the halo, making them oblate or nearly spherical, but allowing them to remain triaxial at intermediate radii and elongated at larger radii \cite{2004ApJ...616...16G,2004ApJ...611L..73K,2006PhRvD..74l3522G,2008ApJ...681.1076D,2010MNRAS.405.2161D,2012ApJ...748...54Z}.
In \cite{Arora22} it has been shown on the basis of cosmological simulations of MW-like galaxies that orbits of stars cannot be reliably reconstructed in static models of potential if the real system experience mergers with 1:8 or higher mass ratio. An alternative considered in \cite{Arora22,2024ApJ...977...23A} is to use models with evolving potential using expansion into some basis functions.




The complex and time dependent shape of galactic potentials predicted by cosmological simulations can be described by basis function expansion. Authors of \cite{2020MNRAS.499.4793S} have shown that with the reasonable accuracy the angular variations of the potential of a MW-like halo can be described by a moderate number (4--10) of spherical harmonics, and the radial variations can be described by splines.   The method used for this expansion is available in the code \texttt{AGAMA} \cite{2018arXiv180208255V}. However, the sample of MW-like halos in \cite{2020MNRAS.499.4793S}, as well as in \cite{Arora22}, lacks an environment similar to that of the Galaxy. It is known that the Galaxy is located in the Local Group (LG) at the edge of the Local Void and there is a Local Filament passing through the LG \cite{Klypin03,Forero-Romero15,Tully13,Tully14,Libeskind15}.  These structures should define the general flow of matter during the (still ongoing) formation of the MW \cite{2011MNRAS.411.1525L}. Besides this, the mass distribution in the LG could itself affect at least the quadrupole component of the Galactic potential, with LG contribution being comparable to that of the LMC (we show this in the Appendix~\ref{sec:app_cl}).

We use the spherical harmonic expansion approach to analyze the behavior of the potential of MW-like galaxies simulated within a realistic large-scale environment in HESTIA project \cite{2020MNRAS.498.2968L}. Our main aim is to check whether the LG mass distribution should be taken into account while modeling the MW potential. Also we search for similarities between the different random realizations of MW-like objects in constrained simulations --- if such similarities exist, they could be expected in the real Galaxy.

The paper is organized as follows: in Section~\ref{sec:axis} {we describe HESTIA simulation suite and general parameters of LG analogues}. In Section~\ref{sec:components} in Subsection~\ref{sec:expansion} we introduce the potential expansion into spherical harmonics and give the details of our usage of \texttt{AGAMA}. In Subsection~\ref{sec:evolution} we describe the time evolution of the gravitational potential. Section~\ref{sec:environment} is devoted to the analysis of the role of the environment in creating the variations of the potential. Results are briefly summarized in Section~\ref{sec:conclusions}. In Appendix~\ref{sec:app_rmax} we give details on the choice of the maximal distance used to compute the potential. In Appendix~\ref{sec:app_cl} we make a simple estimate of LMC impact and LG impact on the non-sphericity of the potential and present the amplitudes of spherical harmonics as a function of distance for all HESTIA realizations. In Appendix~\ref{sec:app_alm} we 
check the effects of numerical and temporal resolution on our results and in Appendix~\ref{sec:all_alm} we show an example of the full set of coefficients for one realization.


\section{Simulations and LG analogues}\label{sec:axis}
Simulations from the HESTIA project are made  in a box with $L_{box}=100$ Mpc/h side based on the reconstruction of initial conditions from the measured distribution of velocities of the galaxies around us. They successfully reproduce the large-scale environment of the Local Group (Virgo cluster, local filament, etc.), and the main parameters of the Local Group itself: the masses of the two main galaxies, their relative distance and velocity. The set of models contains 14 realizations in which other, unmeasured parameters are varied. 13 of these realizations are described in \cite{Libeskind20}, they belong to three groups with identical initial random seeds for the long wave part of the initial conditions, corresponding to $256^3$ mesh in the full box. Within each group simulations have different randomly added small scale structure.

We use two kinds of zoom-in hydrodynamical simulations: the intermediate resolution runs have a spatial and mass resolution corresponding to $4096^3$ particles in the box. This resolution is achieved in a 6 Mpc (4 Mpc/h) radius spherical blob around the center of LG, which is free of low resolution particles. These intermediate resolution simulations are available for all 14 realizations and are further referred as `4k' realizations. From these 14 realizations, three are also simulated with high resolution corresponding to $8192^3$ particles (`8k' simulations) within the region composed from two overlapping spheres around two LG main galaxies with 3.7 Mpc radius each. {Far outside the zoom region the resolution corresponds to $256^3$ particles in the box.} From all these simulations we have considered only the last 38 snapshots which corresponds to the last six Gyr (redshifts $0<z<0.625$). We also have used a simulation for one of the three `8k' realizations where the last billion years is sampled with a high time cadence of 190 snapshots. 
All the distances in this Paper are given in physical (not comoving) units. We use the Planck cosmological model with $H = 100h = 67.7$~km/s/Mpc, $\Omega_m=0.31$, $\Omega_\Lambda=0.69$. {Halos and subhalos were identified using the publicly available \texttt{AHF}\footnote{ \texttt{AHF} is publicly available from \url{http://popia.ft.uam.es/AHF/}} halo finder \cite{Knollmann09}.}

{Unlike \cite{Libeskind20}, where the smallest of the two main halos was selected as the MW analogue, we do this based on their orientation in supergalactic coordinates: we chose MW so that M31 is always in the ($+,-$) quadrant of the (SGX, SGY) coordinate system centered on the MW. The main properties of the two main halos at $z=0$ and $z=0.625$ (approximately 6~Gyr ago) are given in Table \ref{tab:LG}. The realization 01\_12 was not included in the LG sample of \cite{Libeskind20} since it has too massive MW analogue, but we decided to keep it.}


\begin{table*}[]
    \centering
    \caption{Main parameters of HESTIA LG analogues. Physical distances are given.}
    \begin{tabular}{|c|c|c|c|c|c|c|}
    \hline
LG code & MW mass & MW mass & M31 mass & M31 mass & distance & distance \\
 & 6 Gyr ago & $z=0$ & 6 Gyr ago & $z=0$ & 6 Gyr ago & $z=0$ \\
 & $10^{12}$M$_\odot$ & $10^{12}$M$_\odot$ & $10^{12}$M$_\odot$ & $10^{12}$M$_\odot$ & kpc & kpc \\ 
 \hline
1\_12 &   2.686 &   2.933 &   1.938 &   2.532 &    1152 &     724 \\
9\_10 &   0.986 &   1.252 &   1.566 &   2.063 &     909 &     857 \\
9\_16 &   1.169 &   1.254 &   1.719 &   2.074 &     891 &     752 \\
9\_17 &   1.079 &   1.311 &   1.729 &   2.271 &     983 &    1086 \\
9\_18 &   1.224 &   1.919 &   2.155 &   2.225 &     913 &     856 \\
9\_19 &   1.019 &   1.218 &   1.853 &   2.153 &     852 &     726 \\
17\_10 &   1.792 &   2.078 &   1.769 &   2.162 &     979 &     736 \\
17\_11 &   1.600 &   1.983 &   1.772 &   2.347 &     998 &     662 \\
17\_13 &   1.782 &   2.077 &   1.688 &   1.890 &    1043 &     975 \\
17\_14 &   1.914 &   2.200 &   1.731 &   2.064 &     991 &     613 \\
37\_11 &   0.919 &   1.118 &   0.941 &   1.110 &     794 &     864 \\
37\_12 &   0.875 &   1.223 &   0.803 &   0.993 &     832 &     838 \\
37\_16 &   0.797 &   1.083 &   0.801 &   1.113 &     836 &     721 \\
37\_17 &   0.939 &   1.232 &   0.782 &   0.967 &     859 &     805 \\
\hline
    \end{tabular}
    \label{tab:LG}
\end{table*}

\section{Components of Galaxy's gravitational potential} 
\label{sec:components}
\subsection{Potential expansion into spherical harmonics} \label{sec:expansion}
{We introduce a ``Galactic'' coordinate system in each MW analogue by moving the origin of coordinates to the center identified by \texttt{AHF}. We do not rotate this coordinate system, so its orientation is roughly coinciding with the supergalactic coordinate system (i.e. the positions of local structures, such as Virgo cluster, in our coordinate system are close to those in the supergalactic coordinates of the real Universe). One should note that the halo center in \texttt{AHF} is identified as the highest density peak of the gravitationally bound particles. All types of particles (dark matter and baryonic) participate in the center finding.}

{Our reference frame moves with acceleration with respect to the global simulation frame, so it is a non-inertial one, but it resembles the coordinate system used when dealing with observations in our Galaxy. The position of the coordinate system origin also differs from the halo center of mass. Consequently, using our results of potential expansion to compute orbits will require to take into account a fictious force, which can compensate for the reference frame motion, see \cite{2020MNRAS.499.4793S,2024ApJ...977...23A} for the details on its computation.}

Using the {`Multipole' version of gravitational potential constructor of the} \texttt{AGAMA} code we expand the gravitational potential of the simulated MW analogues {(given by sets of particles from N-body snapshots)} into a set of spherical harmonics:
\begin{equation}
    \Phi(r,\theta,\varphi) = \sum_{\ell ,m} a_{\ell m}(r) Y_{\ell m}(\theta,\varphi),
    \label{eq:alm}
\end{equation}
where $a_{\ell m}(r)$ {(multipole expansion coefficients)} are quintic splines interpolating values tabulated on a logarithmic grid in radius $r$ {using $n$ spline nodes}. We use all available particles (DM, stars and gas) from simulations in a range of radii $r_\mathrm{min}<r<r_\mathrm{max}$. The choice of $r_\mathrm{min}$, $r_\mathrm{max}$ is discussed below. The potential is computed by first calculating the spherical harmonics expansion of the density, and then the Poisson equation is solved for the expansion coefficients. 

As is known from literature, halos can demonstrate figure rotation \cite{1992ApJ...401..441D,2004ApJ...616...27B,2007MNRAS.380..657B}, so it is also useful to consider a combination of the coefficients $a_{\ell m}$ which characterizes the amplitude of non-sphericity independently on the orientation:
\begin{equation}
    c_{\ell}^2 = \sum_{-m}^{m} a_{\ell m}^2.
    \label{eq:cl}
\end{equation}

We start by determining the parameters of the expansion: the number of spherical harmonics $\ell_\mathrm{max}$, radial bins $n$, and the spatial extent ($r_\mathrm{min}$, $r_\mathrm{max}$) of the set of particles used to find the potential. Based on the results of \cite{2020MNRAS.499.4793S} and \cite{Arora22} we decide to use $\ell_\mathrm{max}=4$. We chose $r_\mathrm{min}=5$~kpc, since we are not interested in the inner part of the Galaxy and we analyze the potential starting from 10~kpc from the center. Decreasing $r_\mathrm{min}$ to 0.5~kpc does not change our results. To determine $r_\mathrm{max}$ and $n$ we vary these parameters and explore the stability of the results, and find that $r_\mathrm{max}=3$~Mpc and $n=40$ provide the best convergence. We further justify the choice of $r_\mathrm{max}$ in the Appendix~\ref{sec:app_rmax}. Additionally, we show that it is insufficient to limit $r_\mathrm{max}$ to $r_\mathrm{vir}$ for MW-like objects in Appendix~\ref{sec:app_cl} by comparing the impact of a LMC-like object and an M31-like object on the MW potential. We find that these two different disturbers have comparable impact at distances 100-200~kpc.


{We check contributions of baryonic and dark matter to the non-spherical part of the potential. For that, we compute $c_\ell$ created by only baryonic and only DM particles. For all the realizations and $\ell\ge 2$ DM starts to dominate at distances $r>25$~kpc. At $r>50$~kpc the ratio of $c_\ell$ created by DM to the total $c_\ell$ stabilizes at $0.8-0.9$.}
 
{Results of the multipole expansion of gravitational potential of HESTIA MW analogues can be obtained from Github \cite{potential-data}. 

\subsection{Evolution of the gravitational potential} \label{sec:evolution}
Now, after determining the parameters of the potential expansion explained in Section~\ref{sec:expansion} we calculate the evolution of $a_{\ell m}$ for MW analogues. We track the coefficients $a_{\ell m}$ at distances r=(10,20,30,40, 50,100,150)~kpc. These distances are somewhat arbitrary, but 10~kpc is inside the Galactic disk and close to the Solar system distance from center, 20---50~kpc is where the most of tidal streams are now detected by Gaia, and 100---150~kpc are within the Galactic halo. We consider the last 6 Gyr of evolution which corresponds to $z<0.63$. During that time virial radii of MW analogues grow on average by 40\%, but we believe that the central part of the MW is more stable.  That's why we analyze the potential at fixed radii, and not at fixed fractions of the virial radius.

\begin{figure*}
\begin{flushright} 
 \includegraphics[width=0.99\textwidth]{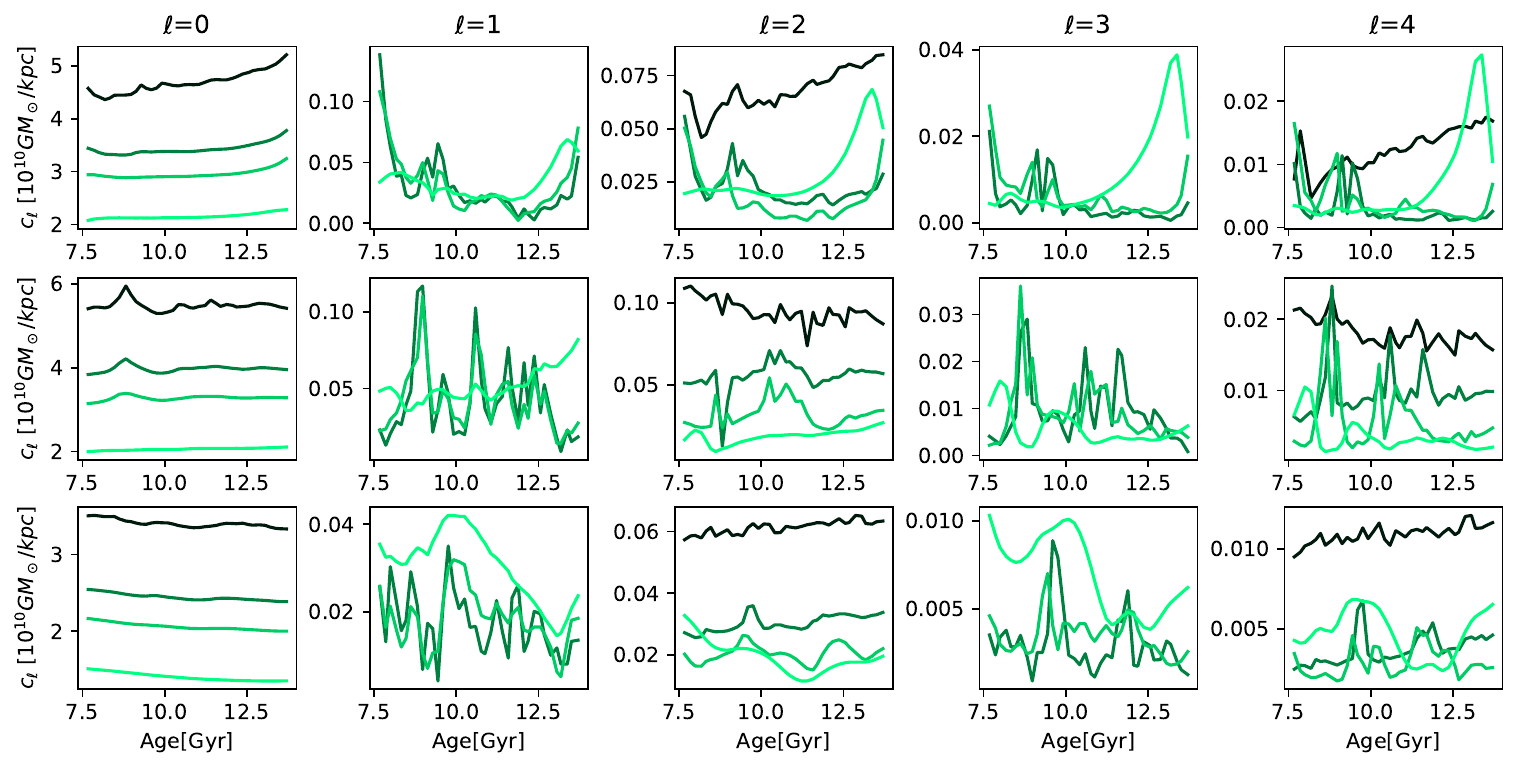}
 \caption{Amplitudes of spherical harmonics as a function of time along four radii. Colors from dark to light correspond to radii of 10, 30, 50 and 150~kpc. Top row: realization 09\_18, middle row: 17\_11, bottom row: 37\_11. {We removed results for 10~kpc radius for odd $\ell$ since they are dominated by noise due to the errors in center determination.}}
 \label{fig:Age_c_l}
\end{flushright}
\end{figure*}

The evolution of the coefficients $c_\ell$ for the three high resolution realizations is shown in Fig.~\ref{fig:Age_c_l}. As has been discussed in Section~\ref{sec:expansion}, these coefficients are not sensitive to the rotations of the potential. An example of the full set of $a_{\ell m}$ coefficients is shown in Appendix~\ref{sec:all_alm}. The convergence of $a_{\ell m}(t)$ between intermediate and high resolutions and with the change of snapshot frequency is discussed in Appendix~\ref{sec:app_alm}. From Fig.~\ref{fig:Age_c_l} it is seen that all three realizations experience significant (more than 30\%) variations of $c_\ell$ with various duration from 0.2 Gyr to several Gyr. The central parts of MW analogues show somewhat more rapid fluctuations. The analysis of the realization with high time cadence of snapshots shows that there are no such significant variations at timescales shorter than 0.2 Gyr.

For the three realizations shown in Fig.~\ref{fig:Age_c_l}, we compare the evolution of $c_\ell$ with the trajectories of the most massive satellites, and find the following:
\begin{enumerate}
    \item For 09\_18 (top row) the highest spike in the lightest curve ($r=150$~kpc) is clearly connected with the accretion of a massive ($4\times10^{11}$~M$_\odot$) satellite which crosses the virial radius one gigayear ago. The spikes in $c_{\ell}$ at 30 and 50~kpc are caused by a less massive satellite ($2\times10^{10}$~M$_\odot$) reaching these distances.
    \item For 17\_11 (middle row) wide peaks at 150~kpc for $\ell=3,4$ and more narrow peaks at 30 and 50~kpc are also caused by passages of a single satellite which was first accreted 6 Gyr ago with a mass of $1.3\times10^{11}$~M$_\odot$ and then it made several pericenter passages, gradually loosing mass.
    \item For 37\_11 (bottom row) a satellite with a mass of {$2\times10^{10}$~M$_\odot$ is orbiting around MW at distances $100-200$~kpc and it is responsible for the large variations of the 150~kpc curve for $\ell=1,3$.}
\end{enumerate}
In almost all the cases, besides $\ell=1$, the variations of coefficients at $r=10$~kpc do not show clear connections with the most massive satellites which do not reach such close distances to the center.

\section{Impact of the environment} \label{sec:environment}

As we have already shown in Section \ref{sec:expansion} and Appendix~\ref{sec:app_cl}, there is a significant impact of the local environment of the Galaxy on its $a_{\ell m}$. Here we analyze this impact in more detail. {To illustrate the effect of the environment on the observable properties of modeled galaxies, we compute trajectories of test particles and measure the change of their angular momenta. Due to Noether's theorem, in a spherically-symmetric potential the angular momentum is conserved, while in a general non-spherical potential all components of the angular momentum are not conserved, therefore the change of the angular momentum can characterize non-sphericity.}

{We initially place test particles on random circular orbits with fixed radius in a monopole component of the measured potential 6~Gyr ago. Then we compute particle trajectories in the time-dependent potential constructed from the measured multipole expansion coefficients up to the present time, taking into account the fictious force due to the reference frame acceleration. We compare the the angular momenta $\mathbf{L}_\mathrm{final}$ after such calculation with the initial one $\mathbf{L}_\mathrm{init}$ (the latter is identical for all the particles on a similar radius). We do such calculations for all the realizations and for two kinds of potential expansion: the first is limited to the virial radius of the MW halo ($r_\mathrm{max}=r_\mathrm{vir}$), and the second takes into account all the matter within 3~Mpc from MW center ($r_\mathrm{max}=3$~Mpc, as proposed in Appendix~\ref{sec:app_rmax}). Fig.~\ref{fig:Lchange} shows the average and scatter of the distributions of the relative change of the angular momentum, $|\mathbf{L}_\mathrm{final}-\mathbf{L}_\mathrm{init}|/|\mathbf{L}_\mathrm{init}|$ for these cases.}

{From Fig.~\ref{fig:Lchange} one can see that for the initial orbit radius of 50~kpc there is no significant difference in the change of the angular momentum when the environment is included. At 100 and 150~kpc the impact is more pronounced: in all the realizations the addition of the environment makes the difference of the angular momentum larger. Note that Fig.~\ref{fig:Lchange} shows the change in the angular momentum vector (we indicate vectors by bold font). We also compare amplitudes of the angular momentum and find that for the initial orbit radius of 100~kpc the mean of $L_\mathrm{final}/L_\mathrm{init}$ is 0.88 and 0.91 for $r_\mathrm{max}=3$~Mpc and $r_\mathrm{max}=r_\mathrm{vir}$ correspondingly, and the RMS scatter of this ratio is 0.31 and 0.23. Angular momenta of individual particles also differ significantly: the RMS scatter of $L_\mathrm{final}(r_\mathrm{max}=3\;\mathrm{Mpc}) / L_\mathrm{final}(r_\mathrm{max}=r_\mathrm{vir})$ is 0.35 (the initial positions and velocity orientations were identical for orbit runs with different $r_\mathrm{max}$).}

{Using the same approach with the measurements of the angular momentum we also verify our choice of $\ell_\mathrm{max}=4$. Specifically, we compute orbits for one realization with $\ell_\mathrm{max}=8$ and find that the difference in the angular momentum $L_\mathrm{final}$ with the $\ell_\mathrm{max}=4$ case is within 5\%.}

\begin{figure*}
    \centering
    \includegraphics[width=\linewidth]{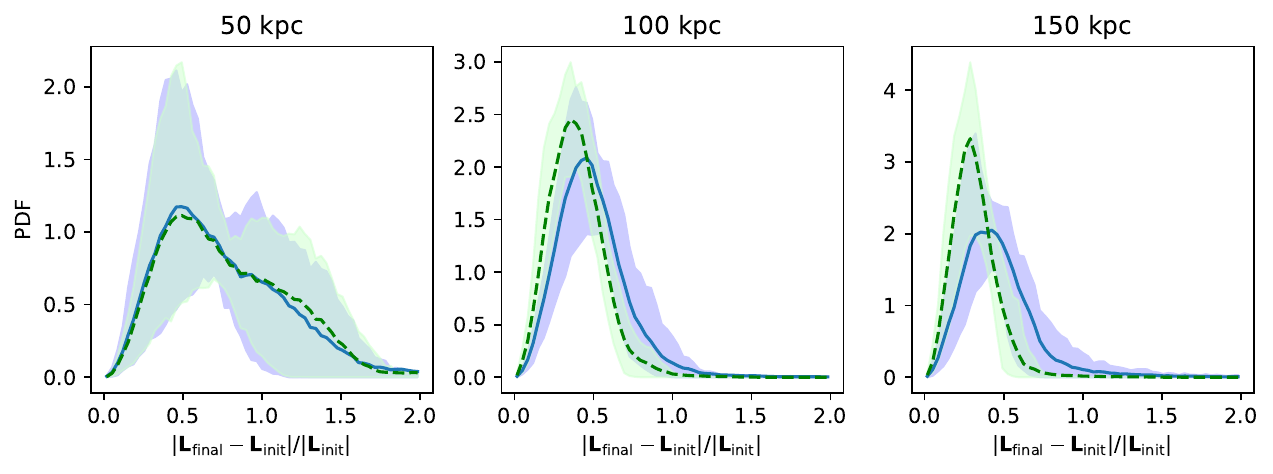}
    \caption{{Distribution of the change in test particle angular momentum after 6~Gyr of evolution for three initial radii (50, 100 and 150~kpc from left to right). Blue solid lines show the mean over realizations for the expansion to $r_\mathrm{max}=3$~Mpc, green dashed lines correspond to $r_\mathrm{max}=r_\mathrm{vir}$. Shaded regions show scatter over realizations from 10\% to 90\%.}}
    \label{fig:Lchange}
\end{figure*}

{Next, we check if the non-sphericity of the potential is created by nearby galaxies, or the diffuse matter also plays a role.} For that we compare the potential computed for the simulation particles with the potential computed for all nearby galaxies and halos except for the MW itself (but including the MW satellites), considered as point masses. We demonstrate this comparison in Fig.~\ref{fig:Age_a_lm_pot_gal}. This figure shows $c_{\ell}(t)$ at three radii for the realization of 09\_18 with a resolution of `8k'. {We show only $\ell=2$ and $\ell=4$, but the multipoles with $\ell=1,3$ show qualitatively similar behavior.} There is a clear correlation between potential evolution computed from particles and from nearby galaxies. This correlation is weaker for $r=10$~kpc (darkest lines in Fig.~\ref{fig:Age_a_lm_pot_gal}). We expect this, since the potential at $r=10$~kpc should have high contribution of the galactic disk, which is not traced by nearby galaxies. This is illustrated in Fig.~\ref{fig:mtot-mgal}, where the ratio of the total mass in particles and in nearby galaxies is shown. At $r\sim10$~kpc the mass in particles is 5-9 times higher than in galaxies, while at larger distances this ratio converges to 2-2.5.

To quantify the correlation we calculate the coefficients:
\begin{equation}
    \xi_{\ell m} = \frac{\langle a_{\ell m}^\mathrm{full} \; a_{\ell m}^\mathrm{gal} \rangle }{\mathrm{\sigma} (a_{\ell m}^\mathrm{full}) \; \mathrm{\sigma} (a_{\ell m}^\mathrm{gal})},
    \label{eq:xi}
\end{equation}
where $a_{\ell m}^\mathrm{full}$ is computed for the full particle distribution, $a_{\ell m}^\mathrm{gal}$ is for nearby galaxies considered as point masses, $\sigma$ denotes the standard deviation.


\begin{figure*}
\begin{center}
 \includegraphics[width=0.49\textwidth]{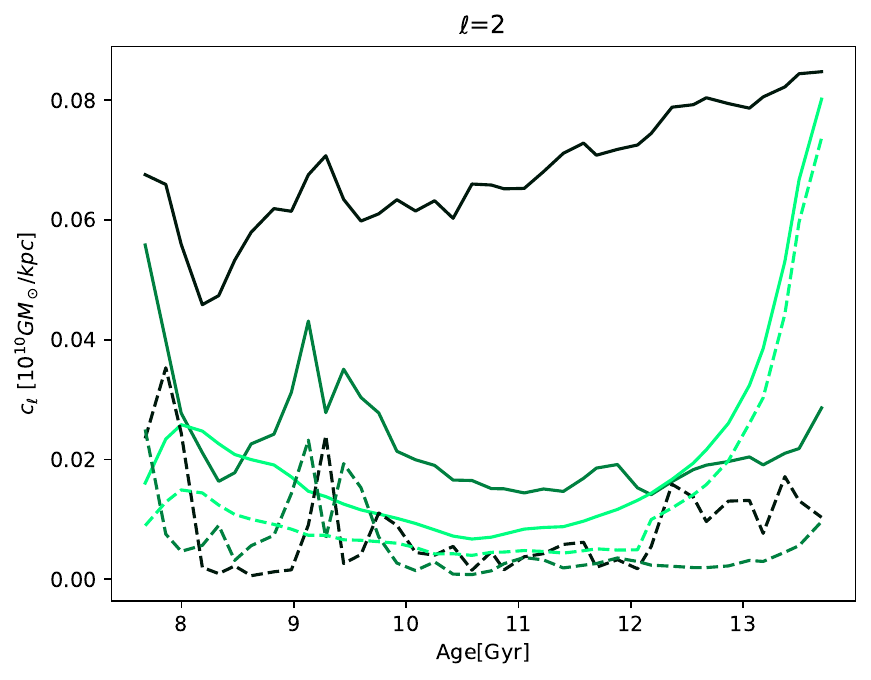} 
 \includegraphics[width=0.49\textwidth]{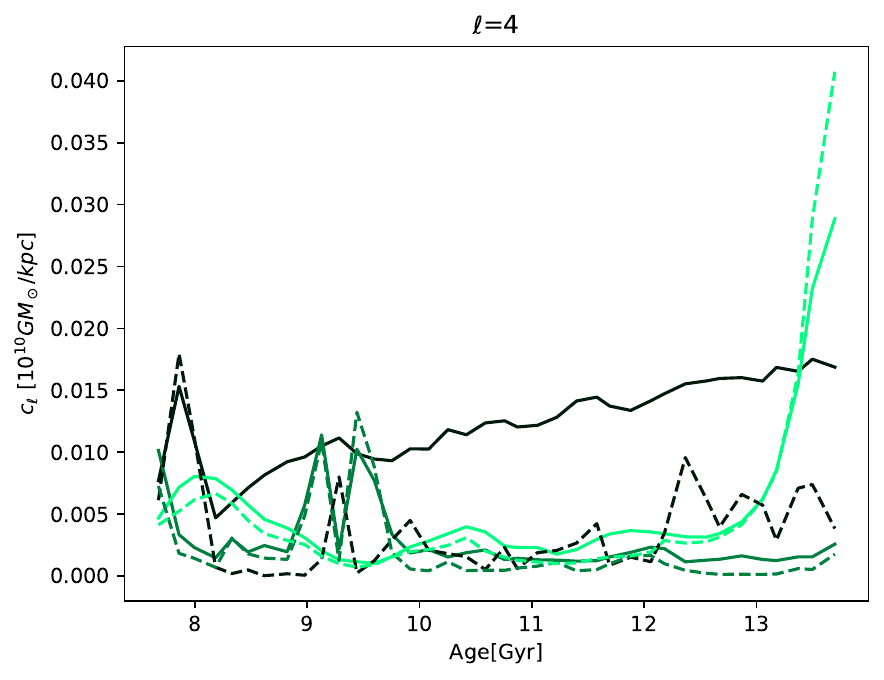} 
 \caption{Impact of MW satellite galaxies and LG galaxies on the evolution of the potential harmonics $c_{2}$ (left panel) and $c_{4}$ (right panel). The lines are plotted for 10, 30 and 100 kpc (in order of increasing line brightness). Solid lines are calculated for the full mass distribution while dashed lines are for the satellites and surrounding galaxies and halos only.}
 \label{fig:Age_a_lm_pot_gal}
\end{center}
\end{figure*}

\begin{figure}
    \centering
    \includegraphics[width=\linewidth]{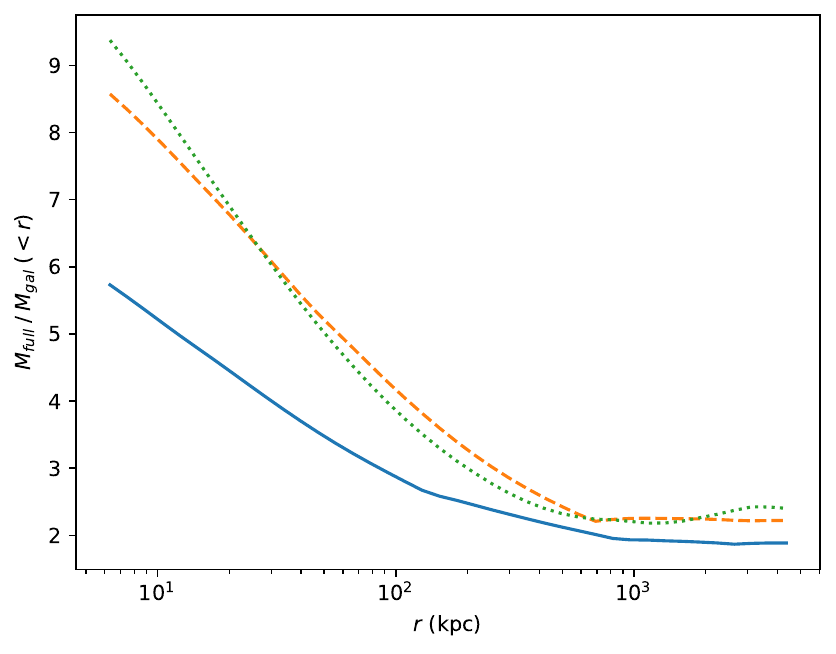}
    \caption{Ratio of the total mass in simulation particles to the mass in nearby galaxies from distance to MW center for the three high resolution realization: 09\_18 (solid), 17\_11 (dashed) and 37\_11 (dotted).}
    \label{fig:mtot-mgal}
\end{figure}


Over all `8k' realizations and snapshots the correlation between $a_{\ell m}$ computed from the nearby galaxy positions and from the total mass distribution for $\ell>1$ is $\xi=0.87$ for 150~kpc, $\xi=0.83$ for 100~kpc and $\xi=0.77$ for 50~kpc. 

We point again that it is not sufficient to consider only the MW satellites when computing the evolution of the potential. If we construct the potential using only galaxies within the virial radius, the potential expansion coefficients in `8k' simulations show no significant correlation with $a_{\ell m}$ for the total mass distribution, while in `4k' simulations (which have more realizations) there is a correlation on the level of $\xi=0.68$ for all harmonics with $\ell>1$ at 100~kpc.

We also investigate the correlation of the potential shape with the position of the Andromeda galaxy, for which we {find the expansion coefficients for a single point mass located at the Andromeda position} and compute the correlation of these values with $a_{\ell m}$ over all snapshots and realizations. We find a correlation coefficient above 0.6 only for the quadrupole at 150~kpc. This means that other members of the LG are also need to be taken into account when modeling the MW potential.

The relatively high values of correlations $\xi$ between $a_{\ell m}^\mathrm{gal}(t)$ and $a_{\ell m}^\mathrm{full}(t)$ at $r>30$~kpc indicate that it might be possible to reconstruct the potential $a_{\ell m}^\mathrm{full}(t)$ from the spatial distribution and masses of nearby galaxies. We try to do it by simply setting $a_{\ell m}^\mathrm{model}(t) = A a_{\ell m}^\mathrm{gal}(t)$, where $A$ is a constant. We measure the quality of reconstruction for each $\ell,m,r$ by computing 
\begin{equation}
    Q = \frac{\mathrm{\sigma}(a_{\ell m}^\mathrm{model} - a_{\ell m}^\mathrm{full})}{\mathrm{\sigma}(a_{\ell m}^\mathrm{full})},
\end{equation}
where the standard deviation is computed across all snapshots and realizations. We vary the constant $A$ in the range 0.5 --- 3.0, but the quality $Q$ never reaches below 0.45 for any of $\ell,m,r$. For $A=1$ we get $Q$ in the range from 0.5 to 1.1 for $\ell>1$, $r\ge 30$~kpc. This means that despite the good correlation between $a_{\ell m}^\mathrm{gal}(t)$ and $a_{\ell m}^\mathrm{full}(t)$, they are not equal to each other. {Note also that we do not take into account the zone of avoidance which affects the distribution of known galaxies in observations.}

We also check for similarities between the potential evolution tracks of different LG realizations. 
{Such similarities are the manifestation of the constraining power of the constraints imposed on the simulations, in the $\Lambda$CDM context. The smaller is the variance between the different realizations the more constrained is the LG.}
As a result, e.g., M31 positions at $z=0$ are not completely random in these simulations, see Fig.~\ref{fig:m31-pos}. To check wether there is a trace of these similarities in the MW potential, we compute the average and scatter of $a_{\ell m}$ across realizations for each simulation snapshot at $r=100$~kpc. We find that {the scatter between $a_{\ell m}$ in different realizations is greater than the mean values of $a_{\ell m}$ across realizations}, so there is no clear similarities between potentials in different realizations. However, {one can see some similarities in Fig.~\ref{fig:all_cl} where for most of the realizations the growth of $c_\ell(r)$ at distances 200---700~kpc is observed. This is an indication that the virial scale is less constrained compared to the quasi-linear scale of the LG as a whole.}



\begin{figure}
    \centering
    \includegraphics[width=\linewidth]{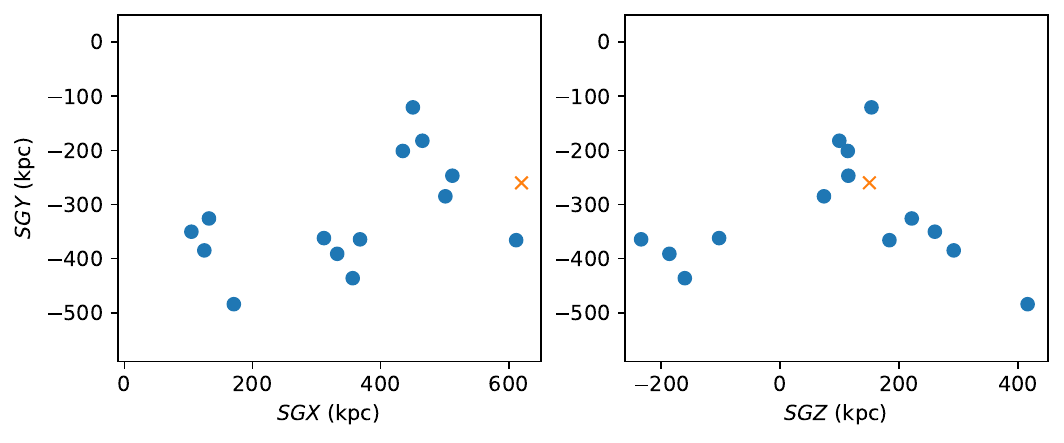}
    \caption{Positions of M31 analogs in 14 LG realizations (circles) and the real M31 (crosses) in supergalactic coordinates.}
    \label{fig:m31-pos}
\end{figure}



\section{Conclusions}\label{sec:conclusions}
In this paper we use constrained simulations from the HESTIA suite to investigate the gravitational potential of MW analogues. Our aim is to quantify the variations of the potential with time and the role of the environment (i.e. the Local Group) in these variations. Based on that it should be possible to predict what could be the variations for our real Galaxy. We use the code \texttt{AGAMA} to calculate expansion of the gravitational potential into spherical harmonics $a_{\ell m}(r)$ over the last 6 Gyr of cosmic evolution (at $z<0.63$). Based on the results of a similar study made for DM halos in a un-constrained simulation \cite{2020MNRAS.499.4793S}, we limit the expansion to $\ell_\mathrm{max}=4$. 


A simple estimate shows that the non-spherical part of the gravitational potential created by the LG can be comparable to that created by the LMC at distances $100-200$~kpc. We attribute the rise of $c_\ell(r)$ with $r$ for $r_\mathrm{vir}<r<500$~kpc in Fig.~\ref{fig:LMC} and Fig.~\ref{fig:all_cl} to the impact of the LG. Also the analysis of the convergence of $c_\ell(r)$ for $r>50$~kpc as a function of $r_\mathrm{max}$, the maximal distance used to compute potential, shows that it is important to take into account the environment at distances to at least 1~Mpc from the MW center (see Fig.~\ref{fig:R_a_lm}). Ignoring the mass distribution outside the virial radius of the MW can result in the errors of the $a_{\ell m}$ of more than 20\% at $r\sim150$~kpc.

Harmonics of the potential expansion change significantly with time. The clearly visible spikes on the $c_\ell(r,t)$ plots (Fig.~\ref{fig:Age_c_l}) correspond to massive satellites crossing radius $r$. The time evolution of the harmonics is different in different LG realizations, but at $r\ge 30$~kpc it correlates well with the evolution of the potential harmonics reconstructed from the masses and positions of MW satellite galaxies and LG members. This is quantified by the correlation coefficient $\xi$ defined in (\ref{eq:xi}) which reaches 0.9. In other words, the tidal forces at $r\ge 30$~kpc are more significant than the shape of the MW halo and the disk (without the satellites). Again, using only satellite galaxies and ignoring the LG reduces the correlation coefficient to 0.7.

We have checked if it is possible to use this correlation to build the MW potential model from positions and masses of surrounding galaxies. The resulting model has deviations from the potential created by the full mass distribution ranging from 50\% to 100\% for different harmonics $a_{\ell m}$.



As a result of this work, we believe that in order to have an accurate model of the evolving Galactic potential it is important to include not only massive satellites, but also galaxies of the Local Group. This could be important for the analysis of the planes of satellites \cite{Kroupa05,Pawlowski12,Arakelyan18}, the globular cluster zone of avoidance \cite{Nikiforov17}, the problem of satellite's apex \cite{Makarov23} and other problems connected with the shape of the Galactic potential.

\begin{acknowledgments}
Authors are grateful to Eugene Vasiliev and Alexander Knebe for useful discussions. Authors are also thank the referees for their comments.

YH is partially supported by the Israel Science Foundation (ISF
1450/24).

The authors gratefully acknowledge the Gauss
Centre for Supercomputing e.V. (\url{www.gauss-centre.eu}) for funding
HESTIA project by providing computing time on the GCS Supercomputer
SuperMUC at Leibniz Supercomputing Centre (\url{www.lrz.de})
\end{acknowledgments}

\appendix

\section{Choice of $r_\mathrm{max}$}
\label{sec:app_rmax}
We explore the behavior of $c_{\ell}$ at two finite radii, 50~kpc and 150~kpc, as functions of the maximal distance. This behavior is shown in Fig.~\ref{fig:R_a_lm} for one of the realizations, 09\_18 (which has the virial radius of 263~kpc). The dipole harmonic vanishes if the center of mass coincides with the coordinate system origin, so the change of dipole in Fig.~\ref{fig:R_a_lm} is expected and reflects the dependency of the center of mass position from the sphere radius $r_\mathrm{max}$.

One can see from Fig.~\ref{fig:R_a_lm} that the quadrupole at $r=150$~kpc does not converge when $r_\mathrm{max}$ reaches the virial radius (if $r_\mathrm{max}=r_\mathrm{vir}$, the difference of $c_\ell$ from the converged value is 20\%). For higher multipoles, $\ell=3,4$, and for the potential at 50~kpc, this effect is weaker. 
From Fig.~\ref{fig:R_a_lm} we conclude that there is a significant impact of the mass distribution up to 0.7~Mpc on the shape of the Galactic potential, which is close to the distance to the M31 galaxy, which is also shown in Fig.~\ref{fig:R_a_lm}. In other words there is some significant tidal force created by the environment. Also we conclude that $c_{\ell}$ does not change with $r_\mathrm{max}$ when $1$~Mpc$<r_\mathrm{max}<10$~Mpc, so we fix our choice to the middle of this interval.

\begin{figure*}
\begin{center}
 \includegraphics[width=0.49\textwidth]{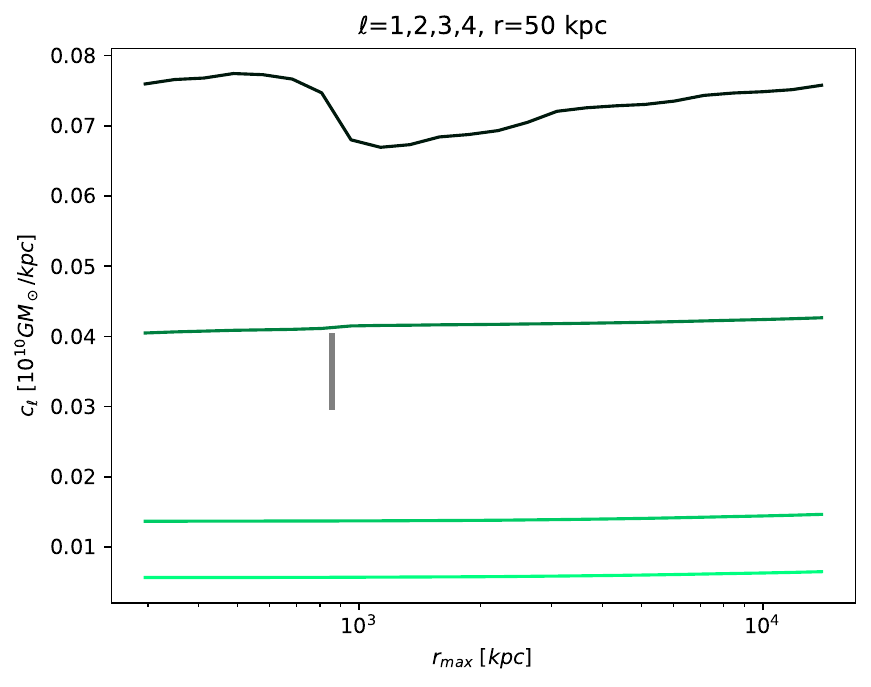} 
 \includegraphics[width=0.49\textwidth]{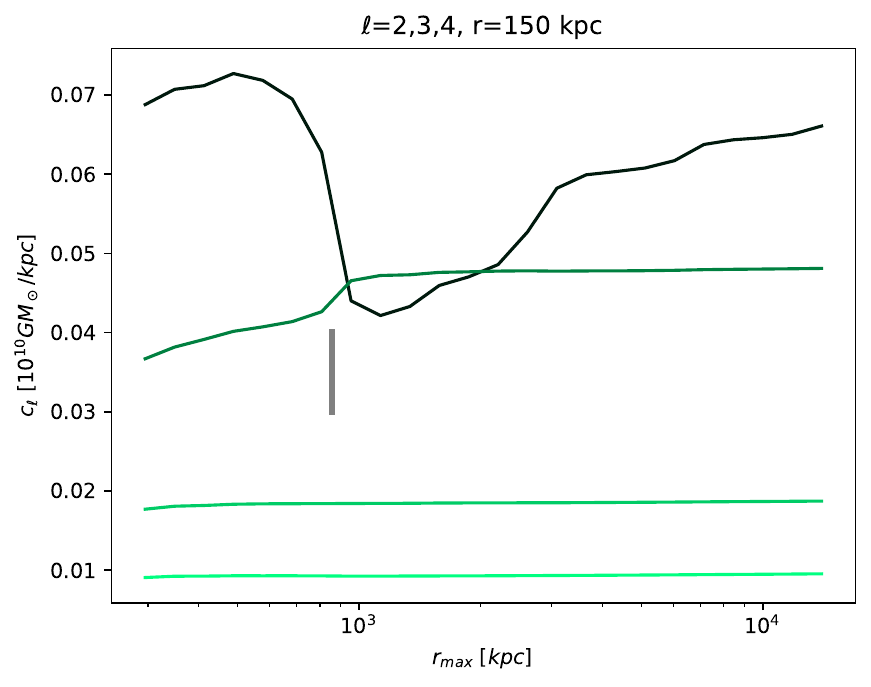} 
 \caption{Amplitude of spherical harmonics for the realization 09\_18 (4k) as a function of maximal distance used to compute the potential. Left panel: computed for $r=50$~kpc, right panel: for $r=150$~kpc. The darkest line shows $\ell=1$, and the lightest line shows $\ell=4$. Vertical line is the M31 analogue distance.}
 \label{fig:R_a_lm}
\end{center}
\end{figure*}

We illustrate the possible impact of the environment by a simple model. Consider a point mass of M31 scale $M_{M31}\sim2\times10^{12}$ M$_\odot$ located at a distance of $d\sim750$~kpc from the center of the MW. For radii $r<d$ coefficients (\ref{eq:cl}) {behave as }
\begin{equation}
\label{eq:cl-r}
c_\ell\propto M\frac{r^\ell}{d^{2\ell+2}},    
\end{equation}
{for $\ell>0$. }
The impact of such a point mass on the non-spherical part of the potential can be compared by an order of magnitude with the impact of the LMC. For that we introduce another point mass $M_{LMC}=2\times10^{11}$~M$_\odot$ at a distance of 50~kpc. The coefficients $c_\ell(r)$ produced by these two point masses are shown in the left panel of Fig.~\ref{fig:LMC}. From this Figure one can conclude that the impact of the M31-like point mass is more significant than that of the LMC at $r>140$~kpc for $\ell=1$, $r>200$~kpc for $\ell=2$ and $r>300$~kpc for $\ell=3,4$. {This motivates us to chose $r_\mathrm{max}>1$~Mpc. Considering the effect of e.g. Virgo cluster (with $M=1.2\times10^{15}$~M$_\odot$ at 20 Mpc) using formula (\ref{eq:cl-r}) shows it should be negligible already for $\ell=1$, this justifies our choice of $r_\mathrm{max}=3$~Mpc.}


\section{Potential non-sphericity for 14 LG realizations}
\label{sec:app_cl}

\begin{figure*}
\begin{center}
 \includegraphics[width=0.49\textwidth]{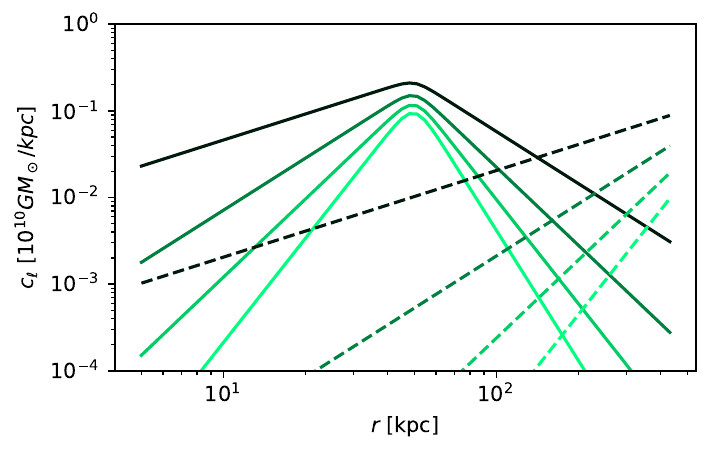}
 \includegraphics[width=0.49\textwidth]{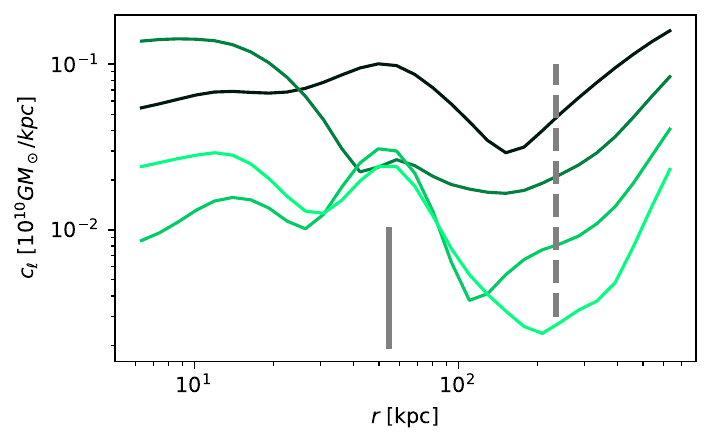}
 \caption{Left panel: Contributions to harmonics with $\ell=1,2,3,4$ (in the order of increasing line brightness) by a LMC-like point mass (solid lines) and M31-like point mass (dashed lines). Right panel: example of potential harmonics for realization 17\_10 at z=0.38 when a satellite with mass $4.9\times10^{10}$~M$_\odot$ is located at 50~kpc from the MW center. Vertical solid line is the satellite position, dashed line is the virial radius.}
 \label{fig:LMC}
\end{center}
\end{figure*}

{The simple estimate presented in Appendix~\ref{sec:app_rmax} gives} the qualitative illustration of the expected effect of the environment {by considering LMC and M31 as point masses}. An example of $c_\ell(r)$ behavior from simulations is shown in the right panel of Fig.~\ref{fig:LMC}. In that particular realization, 17\_10, a satellite with mass $4.9\times10^{10}$~M$_\odot$ flies at 50~kpc from the MW center at $z=0.38$. This satellite produces a clearly seen bell-shaped feature on $c_\ell(r)$ for $\ell=1,3,4$. At distances $r>100$~kpc the amplitude of spherical harmonics starts to rise with radius, what can be a result of the impact of M31 or other masses within LG. This rise is seen for almost all the realizations, see Fig.~\ref{fig:all_cl}. 
The exceptions are the cases where there is a large infalling satellite close to the virial radius.

This is where the constrained simulations of the Local Group can become handy. If there is no impact of the environment, $a_{\ell m}(r)$ will converge for any $r$ when $r_\mathrm{max}\gtrsim r_\mathrm{vir}$. If the environment around MW is affecting its gravitational potential, $a_{\ell m}$ will continue to change with $r_\mathrm{max}$ for $r_\mathrm{max} \gg r_\mathrm{vir}$. This change will depend on the particular mass distribution of the environment, so for the study of MW potential in simulations one needs the environment which resembles the real MW environment.

In Fig.~\ref{fig:all_cl} we show the radial dependency of the amplitudes of spherical harmonics expansion of the potential for all the HESTIA realizations with `4k' resolution. 
From Fig.~\ref{fig:all_cl} one can see that almost all the realizations demonstrate the monotonic growth of $c_\ell(r)$ in the range $200<r<700$~kpc. This is expected, since all HESTIA realizations include a second large halo (M31 analogue) in the LG analogue with the distance from MW in the range $600-1100$~kpc.

{There were no constraints on the LMC analogue in the HESTIA runs, however, so there is a large scatter in positions and masses of the largest subhalos in the MW analogues. This results in randomly positioned bumps in $c_\ell(r)$ functions in different realizations.}
\begin{figure*}
\begin{center}
 \includegraphics[width=0.49\textwidth]{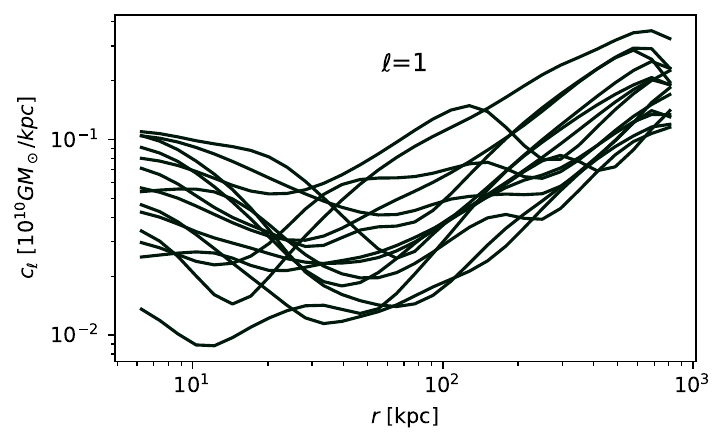} 
 \includegraphics[width=0.49\textwidth]{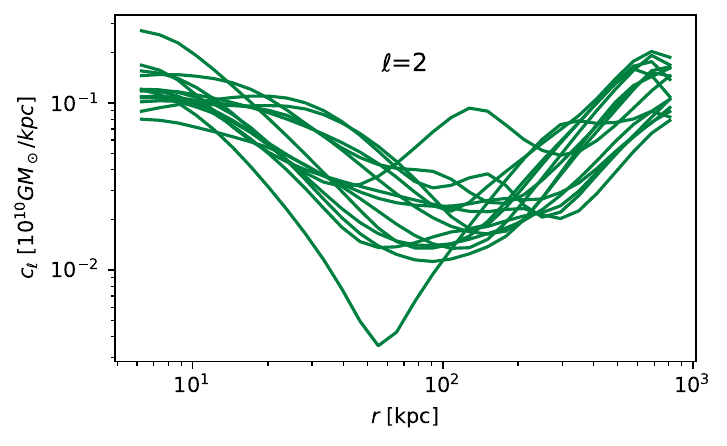} 

  \includegraphics[width=0.49\textwidth]{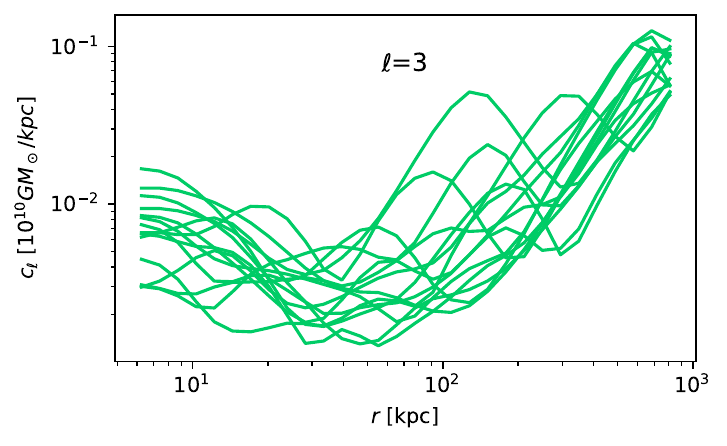} 
 \includegraphics[width=0.49\textwidth]{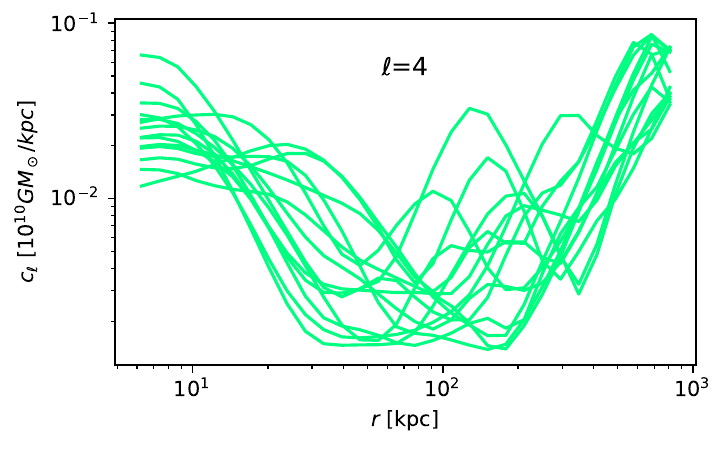} 
 \caption{Coefficients $c_\ell$ introduced in (\ref{eq:cl}) as a function of radius for 14 4k HESTIA realizations at $z=0$. The color shades from dark to light correspond to $\ell=1,2,3,4$.}
 \label{fig:all_cl}
\end{center}
\end{figure*}

\section{Impact of spatial and temporal resolution}
\label{sec:app_alm}

\begin{figure*}
\begin{center}
 \includegraphics[width=0.49\textwidth]{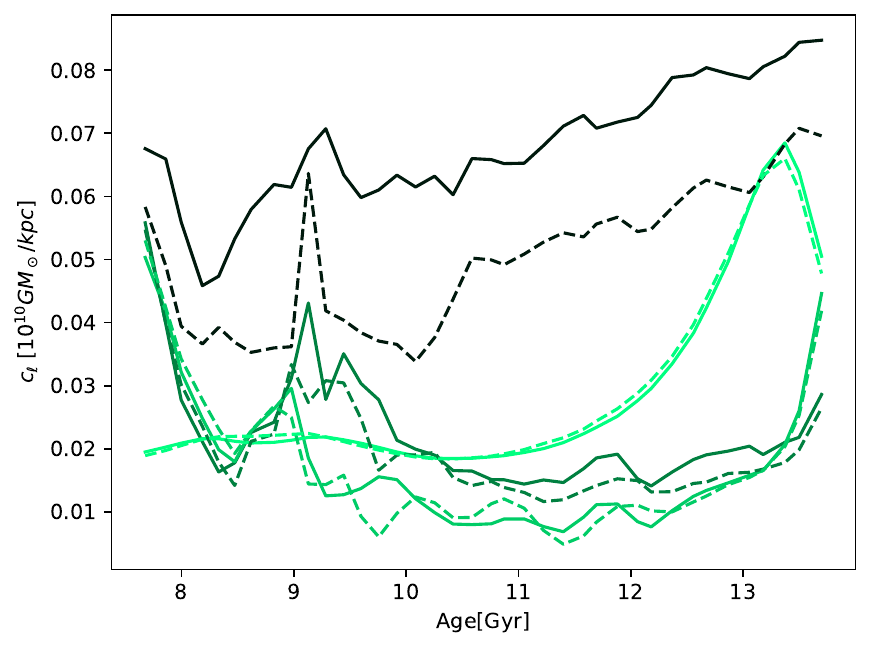} 
 \includegraphics[width=0.49\textwidth]{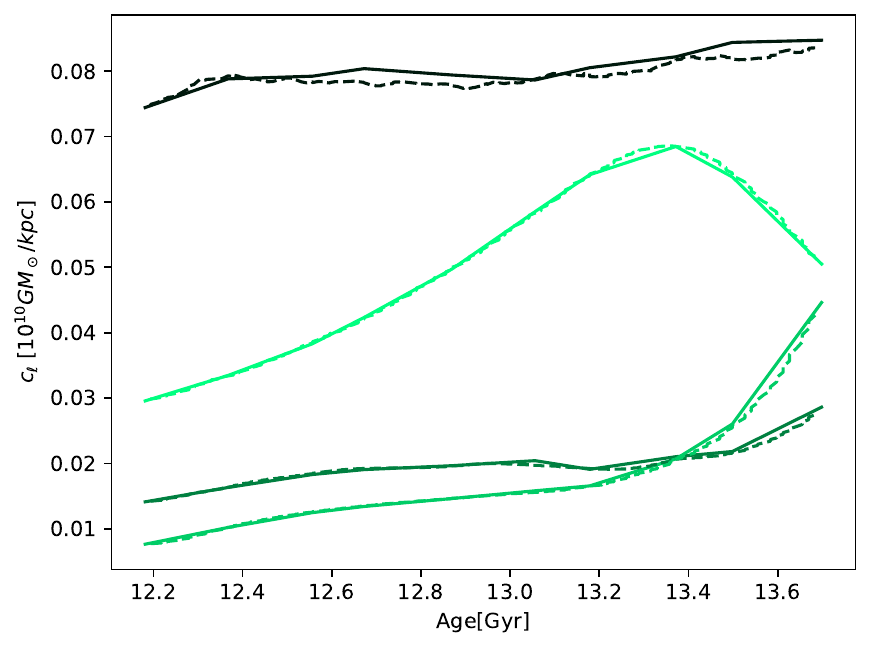} 
 \caption{Expansion coefficients $c_{\ell=2}(t,r)$ of the potential as a function of time are shown for realization 09\_18. Colors from dark to light correspond to radii of 10, 30, 50 and 150 kpc. On the left panel the solid line corresponds to `8k' high resolution run, while the dashed line is for the intermediate resolution `4k' run. On the right panel the solid line represents `8k' run with the normal snapshot frequency and dashed line is for $\sim19$ times higher frequency.}
 \label{fig:Age_a_lm_4k_8k}
\end{center}
\end{figure*}

The HESTIA suite contains simulations with different mass resolution and different snapshot frequency for some of the realizations. This allows us to test the stability and convergence of the $a_{lm}$ evolution measurements. 
In Fig.~\ref{fig:Age_a_lm_4k_8k}, left panel, we compare the evolution of $c_{2}(t,r)$ for realization 09\_18 with zoom resolution of `4k' and `8k'. From this panel one can see that the `4k' simulation has very good agreement with the 8k simulation for all the radii except the 10~kpc. From this we conclude that we reach numerical convergence for $r>10$~kpc for the `4k' simulations.

On the right panel of Fig.~\ref{fig:Age_a_lm_4k_8k} we check if the snapshot frequency of the HESTIA simulations is sufficient to track the change of the potential with time. The data are again for the 09\_18 realization simulated with `8k' resolution. The solid lines show `normal' pace of snapshots (10 snapshots for the last 1.6 billion years), while dashed lines show a simulation with the more frequent pace (190 snapshots). One can conclude from this panel that for all the radii we resolve all the fluctuations of the potential with the `normal' snapshot frequency. 

{We also test the convergence of the `8k' simulations by altering the determination of the halo center which, to our understanding, is the main source of noise for low radii and odd multipoles. For this test we move the origin of the coordinate system to the center of mass of all baryonic particles within 10~kpc from the old center (identified by \texttt{AHF}). Our results show that the multipole expansion coefficients differ by less than 10\% for even harmonics at $r\ge10$~kpc and for odd harmonics at $r\ge20$~kpc, while for odd harmonics at $r=10$~kpc the difference can exceed 50\%.}

\section{Full set of coefficients example}
\label{sec:all_alm}
Fig.~\ref{fig:Age_a_lm_1} and Fig.~\ref{fig:Age_a_lm_2}  show an example of the expansion of the potential in spherical harmonics as a function of time along four radii for realization 09\_18 with a resolution of `8k'. This figure should be compared with Fig.~\ref{fig:Age_c_l}, top panel, where the coefficients are averaged over $m$.

In Fig.~\ref{fig:Age_a_lm_1} and Fig.~\ref{fig:Age_a_lm_2} one can see that some of the $a_{\ell m}$ coefficients demonstrate trends which cannot be seen in Fig.~\ref{fig:Age_c_l}, e.g. $a_{2m}$ and $a_{4m}$ coefficients for $r=10$~kpc demonstrate long-term trends. We expect that the potential at this distance is mostly connected with the disk component and the observed trends may indicate the change of the orientation of the disk rotation axis.

\begin{figure*}
\begin{flushright} 
 \includegraphics[width=1\textwidth]{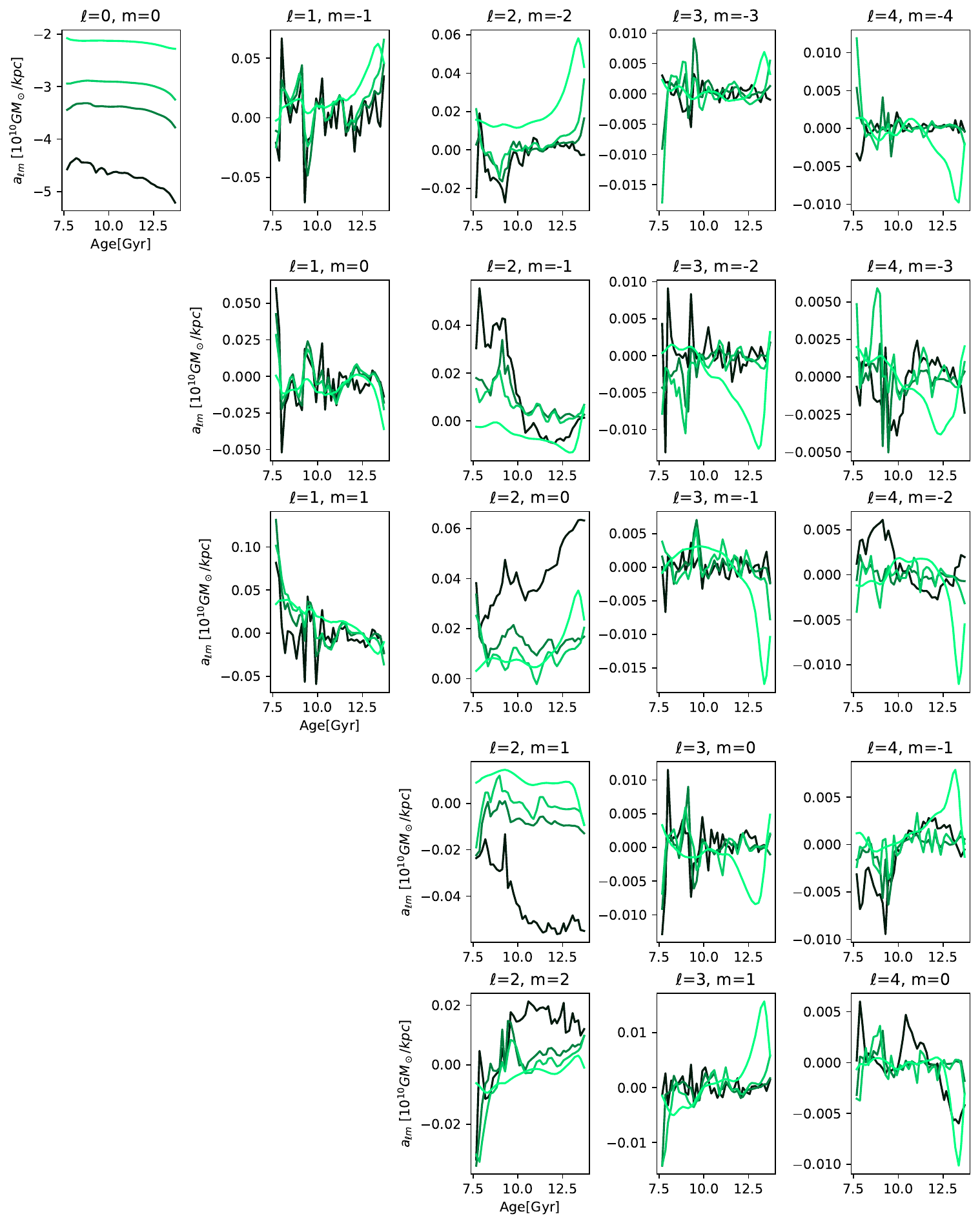}
 \caption{Potential expansion coefficients $a_{\ell m}(t,r)$. Colors from dark to light correspond to $r=10$, 30, 50 and 150~kpc.}
 \label{fig:Age_a_lm_1}
\end{flushright}
\end{figure*}

\begin{figure*}
\begin{flushright} 
 \includegraphics[width=0.45\textwidth]{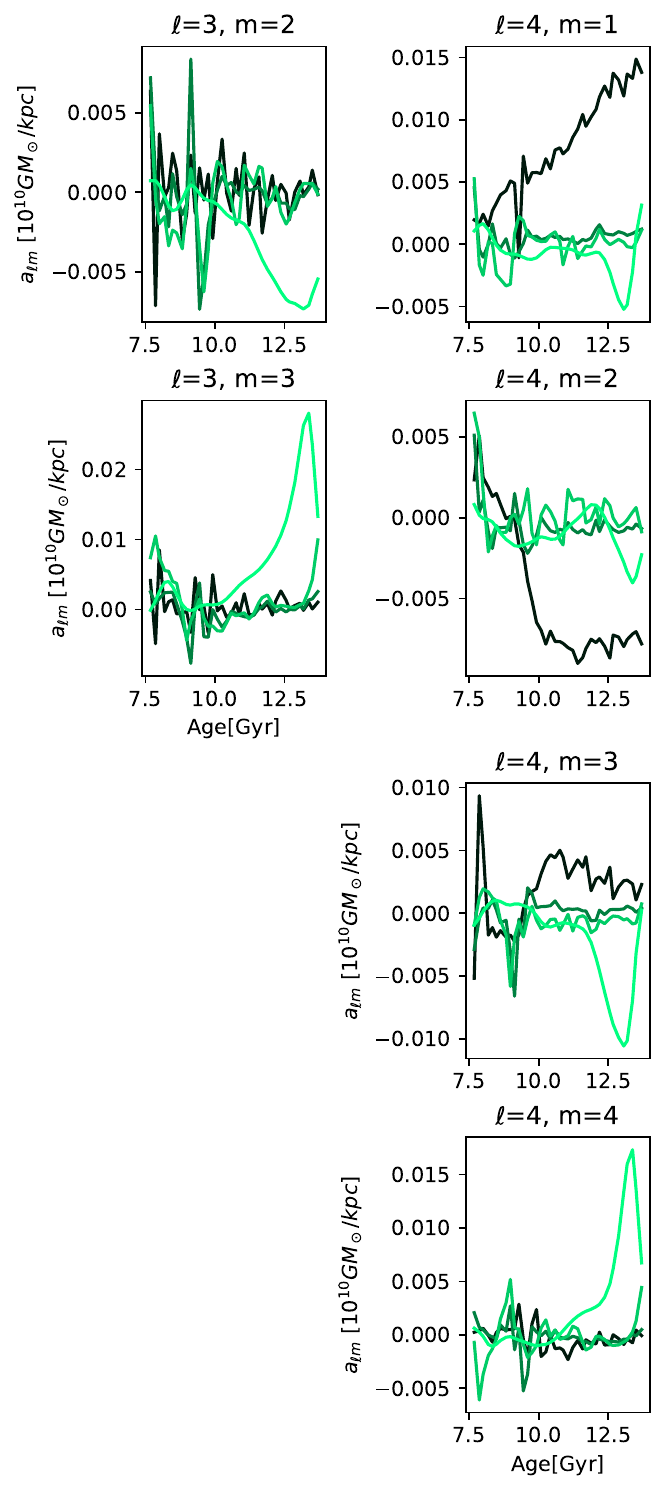}
 \caption{Potential expansion coefficients $a_{\ell m}(t,r)$ (continued). Colors from dark to light correspond to $r=10$, 30, 50 and 150~kpc.}
 \label{fig:Age_a_lm_2}
\end{flushright}
\end{figure*}


 \bibliographystyle{elsarticle-num} 
 \bibliography{Grav_pot}





\end{document}